# Quantum-mechanical nonequivalence of metrics of centrally symmetric uncharged gravitational field

M.V. Gorbatenko, V.P. Neznamov[1]

RFNC-VNIIEF, 37 Mira Ave., Sarov, 607188, Russia

Abstract

Quantum-mechanical analysis shows that the metrics of a centrally symmetric uncharged gravitational field, which are exact solutions of the general relativity equations, are physically non-equivalent.

The classical Schwarzschield metric and the Schwarzschild metrics in isotropic and harmonic coordinates provide for the existence of stationary bound states of Dirac particles with a real energy spectrum. The Hilbert condition $g_{00} > 0$ is responsible for zero values of the wave functions under the "event horizon" that leads to the absence of Hawking radiation.

For the Eddington-Finkelstein and Painlevé-Gullstrand metrics, stationary bound states of spin-half particles cannot exist because Dirac Hamiltonians are non-Hermitian. For these metrics, the condition $g_{00} > 0$ also leads to the absence of Hawking evaporation.

For the Finkelstein-Lemaitre and Kruskal metrics, Dirac Hamiltonians are explicitly time-dependent, and stationary bound states of spin-half particles cannot exist for them. The Hilbert condition for these metrics does not place any constraints on the domains of the wave functions. Hawking evaporation of black holes is possible in this case.

The results can lead to revisiting some concepts of the standard cosmological model related to the evolution of the universe and interaction of collapsars with surrounding matter.

---

[1] E-mail: neznamov@vniief.ru

## 1. Introduction

The Schwarzschild metric [1] is a widely known solution of general relativity for a point uncharged centrally symmetric gravitational field.

The classical Schwarzschild solution is characterized by a spherically symmetric point source of gravitational field of mass $M$ and "event horizon" (gravitational radius)

$$r_0 = \frac{2GM}{c^2}. \tag{1}$$

In (1), $G$ is the gravitational constant, and $c$ is the speed of light. In the classical case, as seen by a distant observer, a test particle reaches the "event horizon" in an infinitely long time.

There are a number of other metrics derived by coordinate transformations of the Schwarzschild metric and also representing exact solutions of general relativity.

We can note the following solutions: the Schwarzschild metric in isotropic coordinates [2], the Schwarzschild metric in harmonic coordinates [3], the Finkelstein-Lemaitre metric [4], the Kruskal metric[2] [5], [6], the Eddington-Finkelstein metric [4], [7], and the Painlevé-Gullstrand metric [8], [9].

In [10] - [12], we developed a method for deriving self-conjugate Dirac Hamiltonians with a flat scalar product of wave functions within the framework of pseudo-Hermitian quantum mechanics for arbitrary, including time-dependent, gravitational fields.

It follows from single-particle quantum mechanics that if the Hamiltonian is Hermitian $\left(\left(\Phi, H\Psi\right) = \left(H\Phi, \Psi\right)\right)$, if there are quadratically integrable wave functions, and if appropriate boundary conditions are specified, the time-independent self-conjugate Hamiltonians should provide for the existence of stationary bound states of spin-half particles with a real energy spectrum.

In [13] - [14], we demonstrated the existence of stationary bound states of Dirac particles by numerical calculations of the Dirac equation for the classical Schwarzschild metric [1]. If the Hilbert condition $g_{00} > 0$ [15], [16] is fulfilled, the "event horizon" of a Schwarzschild black hole represents an infinitely high potential barrier than cannot be crossed by spin-half quantum mechanical particles. The wave function of a Dirac particle under the "event horizon" is zero, that precludes Hawking evaporation of such a black hole [17].

---

[2] The Kruskal metric is derived without coordinate transformations of the Schwarzschild metric

In [18], [19], we also demonstrated the existence of stationary bound states of Dirac particles and the absence of evaporation for the Reissner-Nordström [20], [21], Kerr [22] and Kerr-Newman [23] fields.

In this work we explore the possibility of existence of stationary bound states of Dirac particles for other metrics of uncharged centrally symmetric gravitational field [2] - [9].

As a result of the analysis we will establish that the metrics [1] - [9] can be divided into three groups:

1. For the Schwarzschild metrics [1] - [3], it is possible that non-evaporating black holes with stationary bound states of spin-half particles exist.

2. For the Finkelstein-Lemaitre [4] and Kruskal [5], [6] metrics, Dirac Hamiltonians explicitly depend time, and stationary bound states of Dirac particles cannot exist. The metrics do not impose any constraints on the domain of the wave functions, and Hawking evaporation of black holes is therefore possible for these metrics.

3. The Eddington-Finkelstein [4], [7] and Painlevé-Gullstrand [8], [9] metrics represent an intermediate case. Dirac Hamiltonians for these metrics are non-Hermitian, so the existence of stationary bound states of spin-half particles for them is not possible. On the other hand, the Hilbert condition $(g_{00} > 0)$ leads to the fact that, as in the case of the first group, the "event horizon" is an infinitely high potential barrier, and the wave function of a particle under the "event horizon" is zero. For such black holes, Hawking evaporation does not exist.

It is evident that in terms of quantum mechanics, the three groups of metrics are physically nonequivalent, but all of them are solutions of the general relativity equations and can be implemented during the post-inflationary period of the universe's expansion.

**2. Centrally symmetric solutions of the general relativity equations**

Below we will use the system of units $\hbar = c = 1$, signature

$$\eta^{\alpha\beta} = diag[1,-1,-1,-1] \tag{2}$$

and notation $\gamma^\alpha$, $\gamma^{\underline{\alpha}}$ for global and local Dirac matrices, respectively.

As local matrices we use matrices in the Dirac-Pauli representation.

## 2.1 Schwarzschild solution [1]

The coordinates are

$$(t, r, \theta, \varphi). \tag{3}$$

The interval square is

$$ds^2 = \left(1 - \frac{r_0}{r}\right)dt^2 - \frac{dr^2}{\left(1 - \frac{r_0}{r}\right)} - r^2\left[d\theta^2 + \sin^2\theta d\varphi^2\right]. \tag{4}$$

The domain of variable $r$ is

$$r > r_0. \tag{5}$$

The solution (4) with domain (5) is basic at writing other centrally symmetric solutions because the coordinate transformations are determined in relation to this solution.

## 2.2 Schwarzschild solution in isotropic (isothermal) coordinates [2]

The coordinates are

$$(t, R, \theta, \varphi). \tag{6}$$

The coordinate transformation is

$$r = R\left(1 + \frac{r_0}{4R}\right)^2. \tag{7}$$

The interval square is

$$ds^2 = V^2(R)dt^2 - W^2(R)\left\{dR^2 + R^2\left[d\theta^2 + \sin^2\theta d\varphi^2\right]\right\}. \tag{8}$$

Here

$$V = \frac{\left(1 - \frac{r_0}{4R}\right)}{\left(1 + \frac{r_0}{4R}\right)}, \quad W = \left(1 + \frac{r_0}{4R}\right)^2. \tag{9}$$

It follows from (7) that at transformation $r$ is represented into $R$ in double-valued manner

$$R = \frac{1}{2}\left[\left(r - \frac{r_0}{2}\right) \pm \sqrt{r(r - r_0)}\right]. \tag{10}$$

The domain of $R$ is

$$R > \frac{r_0}{4}. \tag{11}$$

## 2.3 Schwarzschild solution in spherical harmonic coordinates [3]

The coordinates are
$$(t, R, \theta, \varphi). \tag{12}$$

The coordinate transformation is
$$r = R + \frac{r_0}{2}, \tag{13}$$

where $r$ - the Schwarzschild radius, $R$ - harmonic radius.

The interval square is
$$ds^2 = \left(\frac{R - \frac{r_0}{2}}{R + \frac{r_0}{2}}\right) dt^2 - \left(\frac{R + \frac{r_0}{2}}{R - \frac{r_0}{2}}\right) dR^2 - \left(R + \frac{r_0}{2}\right)^2 \left[d\theta^2 + \sin^2\theta d\varphi^2\right]. \tag{14}$$

The domain of $R$ is
$$R > \frac{r_0}{2}. \tag{15}$$

## 2.4 Eddington-Finkelstein metric [4], [7]

The coordinates are
$$(T, r, \theta, \varphi). \tag{16}$$

The coordinate transformation is
$$dT = dt + \frac{r_0}{r} \cdot \frac{dr}{\left(1 - \frac{r_0}{r}\right)}. \tag{17}$$

The condition $\dfrac{\partial}{\partial r}\dfrac{\partial T}{\partial t} = \dfrac{\partial}{\partial t}\dfrac{\partial T}{\partial r}$ is fulfilled but transformation (17) is discontinuous at $r = r_0$.

The interval square is
$$ds^2 = \left(1 - \frac{r_0}{r}\right) dT^2 - 2\left(\frac{r_0}{r}\right) dT dr - \left(1 + \frac{r_0}{r}\right) dr^2 - r^2 \left[d\theta^2 + \sin^2\theta d\varphi^2\right]. \tag{18}$$

The domain of $r$ is
$$r > r_0. \tag{19}$$

## 2.5 Painlevé-Gullstrand metric [8], [9]

The coordinates are
$$(T, r, \theta, \varphi). \tag{20}$$

The coordinate transformation is
$$dT = dt - \sqrt{\frac{r_0}{r}} \cdot \frac{dr}{\left(1 - \frac{r_0}{r}\right)}. \tag{21}$$

The transformation (21) is discontinuous at $r = r_0$; the condition $\dfrac{\partial}{\partial r}\dfrac{\partial T}{\partial t} = \dfrac{\partial}{\partial t}\dfrac{\partial T}{\partial r}$ is fulfilled.

The interval square is
$$ds^2 = \left(1 - \frac{r_0}{r}\right) dT^2 - 2\sqrt{\frac{r_0}{r}}\, dT dr - dr^2 - r^2 \left[d\theta^2 + \sin^2\theta d\varphi^2\right]. \tag{22}$$

The domain of $r$ is
$$r > r_0. \tag{23}$$

## 2.6 Finkelstein-Lemaitre metric [4]

The coordinates are
$$(T, R, \theta, \varphi). \tag{24}$$

The coordinate transformation is
$$dT = dt + \frac{dr\sqrt{\frac{r_0}{r}}}{\left(1 - \frac{r_0}{r}\right)}, \qquad dR = dt + \frac{dr}{\left(1 - \frac{r_0}{r}\right)\sqrt{\frac{r_0}{r}}}. \tag{25}$$

The transformation (25) is discontinuous at $r = r_0$.

The interval square is
$$ds^2 = dT^2 - \frac{dR^2}{\left[\frac{3}{2r_0}(R-T)\right]^{2/3}} - \left[\frac{3}{2}(R-T)\right]^{4/3} r_0^{2/3} \left[d\theta^2 + \sin^2\theta d\varphi^2\right]. \tag{26}$$

The domain of $T, R$ is
$$R > T. \tag{27}$$

**2.7 Kruskal metric [5]**

The Kruskal metric is the further development of the Finkelstein-Lemaitre metric in order to construct the complete reference system for field of a point mass. The below solution form in which reference system is synchronous belong to I.D.Novikov [6]. In coordinates $(\tau, R, \theta, \varphi)$

$$ds^2 = d\tau^2 - \left(1 + \frac{R^2}{r_0^2}\right)(1-\cos\chi)^2 dR^2 -$$
$$-\frac{1}{4}r_0^2\left(\frac{R^2}{r_0^2}+1\right)(1-\cos\chi)^2 \left(d\theta^2 + \sin^2\theta d\varphi^2\right), \tag{28}$$

$$\frac{\tau}{r_0} = \frac{1}{2}\left(\frac{R^2}{r_0^2}+1\right)^{3/2}(\pi - \chi + \sin\chi). \tag{29}$$

The Eqs. (28), (29) show that metric depends on radial coordinate $R$ and proper time $\tau$ through a parameter $\eta$.

Kruskal metric describes a dust space in general case i.e. the space in which there is a non-zero energy-momentum tensor. The Kruskal metric is a generalization of solution of Finkelstein-Lemaitre solution (26) and it is derived without the coordinate transformations of initial Schwarzschild metric (4).

---

One can see that at transformation of Schwarzschild metric to the Eddington-Finkelstein solution (18), the Painlevé-Gullstrand solution (22), the Finkelstein-Lemaitre solution (26) the transformations are discontinuous at $r = r_0$.

It is evident that in this case in the right parts of Einstein equation in the energy-momentum tensor components will be appear additional singular terms. Generally speaking for the general relativity equations only the transformations of $C^3$ class [24], [25] with continuous functions and their first, second and third derivatives with respect to space-time coordinates are acceptable.

Thus, at classical level there is already a question about physical equivalence the metrics under consideration.

Nevertheless all metrics in 2.1 – 2.7 are exact solutions of the general relativity and below we will analyze their quantum-mechanical equivalence within the framework of the possibility of existence of bound states of Dirac particles in gravitational fields of the metrics under consideration digressing from their deriving methods.

Below for uniformity we will consider for all metrics a set of space-time coordinates in similar notation

$$(t, r, \theta, \varphi). \tag{30}$$

## 3. Analysis of the possibility of existence of bound states of spin-half particles in a centrally symmetric gravitational field

To begin with, we present self-conjugate Hamiltonians defined in [12] for the Schwarzschild metric [1] and the metrics [2] - [9].

### 3.1 Schwarzschild metric in coordinates $(t, r, \theta, \varphi)$

$$ds^2 = f_s dt^2 - \frac{dr^2}{f_s} - r^2 \left(d\theta^2 + \sin^2\theta d\varphi^2\right),$$

$$f_s = 1 - \frac{r_0}{r}; \tag{31}$$

The self-conjugate Hamiltonian $H_\eta$ is

$$H_\eta = \sqrt{f_s} m \gamma^0 - i \sqrt{f_s} \gamma^0 \left\{ \gamma^1 \sqrt{f_s} \left( \frac{\partial}{\partial r} + \frac{1}{r} \right) + \right.$$
$$\left. + \gamma^2 \frac{1}{r} \left( \frac{\partial}{\partial \theta} + \frac{1}{2} \operatorname{ctg}\theta \right) + \gamma^3 \frac{1}{r \sin\theta} \frac{\partial}{\partial \varphi} \right\} - \frac{i}{2} \gamma^0 \gamma^1 \frac{\partial f_s}{\partial r}. \tag{32}$$

In (31), (32) we mean real values of $f_s > 0$ (Hilbert condition: $g_{00} > 0$).

### 3.2 Schwarzschild metric in isotropic coordinates

$$ds^2 = V^2(r) dt^2 - W^2(r) \left[ dr^2 + r^2 \left(d\theta^2 + \sin^2\theta d\varphi^2\right) \right], \tag{33}$$

where

$$V(r) = \frac{1 - \frac{r_0}{4r}}{1 + \frac{r_0}{4r}}; \quad W(r) = \left(1 + \frac{r_0}{4r}\right)^2. \tag{34}$$

The self-conjugate Hamiltonian is written as

$$H_\eta = \frac{1 - \frac{r_0}{4r}}{1 + \frac{r_0}{4r}} \beta m + \frac{1 - \frac{r_0}{4r}}{\left(1 + \frac{r_0}{4r}\right)^3} \boldsymbol{\alpha}\mathbf{p} - \frac{i}{2} \alpha^k \left( \frac{\partial}{\partial x^k} \frac{1 - \frac{r_0}{4r}}{\left(1 + \frac{r_0}{4r}\right)^3} \right). \tag{35}$$

In (35), $\beta = \gamma^0$, $\alpha^k = \gamma^0\gamma^k$ are Dirac matrices

It is convenient to write Eq. (35) in the Obukhov form [2]:

$$H_\eta = V(r)\beta m + \frac{1}{2}\left(\boldsymbol{\alpha}\mathbf{p}F_{Ob}(r) + F_{Ob}(r)\boldsymbol{\alpha}\mathbf{p}\right), \tag{36}$$

where

$$F_{Ob}(r) = \frac{V(r)}{W(r)} = \frac{1 - \dfrac{r_0}{4r}}{\left(1 + \dfrac{r_0}{4r}\right)^3}. \tag{37}$$

The Hilbert condition ($g_{00} > 0$) for the solution (33) reduces to the condition

$$r > \frac{r_0}{4}. \tag{38}$$

### 3.3 Schwarzschild metric in harmonic coordinates

$$ds^2 = F_g(r)dt^2 - \frac{1}{F_g(r)}dr^2 - \left(1 + \frac{r_0}{2r}\right)^2 r^2\left(d\theta^2 + \sin^2\theta d\varphi^2\right). \tag{39}$$

In (39),

$$F_g(r) = \frac{1 - \dfrac{r_0}{2r}}{1 + \dfrac{r_0}{2r}}. \tag{40}$$

The self-conjugate Hamiltonian $H_\eta$ is

$$H_\eta = \sqrt{F_g}\beta m - i\alpha^1\left[F_g\left(\frac{\partial}{\partial r} + \frac{1}{r}\right) + \frac{r_0}{2r^2\left(1 + \dfrac{r_0}{2r}\right)^2}\right] -$$

$$-i\sqrt{F_g}\frac{1}{1 + \dfrac{r_0}{2r}}\left[\alpha^2\frac{1}{r}\left(\frac{\partial}{\partial\theta} + \frac{1}{2}\mathrm{ctg}\theta\right) + \alpha^3\frac{1}{r\sin\theta}\frac{\partial}{\partial\varphi}\right]. \tag{41}$$

The Hilbert condition for this solution has the following form:

$$r > \frac{r_0}{2}. \tag{42}$$

### 3.4 Eddington-Finkelstein metric

$$ds^2 = f_s dt^2 - 2\frac{r_0}{r}dtdr - f_{E-F}dr^2 - r^2\left(d\theta^2 + \sin^2\theta d\varphi^2\right),$$

$$f_{E-F} = 1 + \frac{r_0}{r}.$$

(43)

The self-conjugate Hamiltonian $H_\eta$ is

$$H_\eta = \frac{\gamma^0 m}{\sqrt{f_{E-F}}} - i\gamma^0\gamma^1 \frac{1}{f_{E-F}}\left(\frac{\partial}{\partial r} + \frac{1}{r} + \frac{r_0}{r^2}\frac{1}{f_{E-F}}\right) -$$
$$-i\gamma^0\gamma^2 \frac{1}{\sqrt{f_{E-F}}}\frac{1}{r}\left(\frac{\partial}{\partial \theta} + \frac{1}{2}\text{ctg}\theta\right) - i\gamma^0\gamma^3 \frac{1}{\sqrt{f_{E-F}}}\frac{1}{r\sin\theta}\frac{\partial}{\partial \varphi} +$$
$$+i\frac{r_0}{r}\frac{1}{f_{E-F}}\left(\frac{\partial}{\partial r} + \frac{1}{r} - \frac{1}{2rf_{E-F}}\right).$$

(44)

For this solution, the condition $g_{00} > 0$ leads to the condition

$$r > r_0.$$

(45)

### 3.5 Painlevé-Gullstrand metric

$$ds^2 = f_s dt^2 - 2\sqrt{\frac{r_0}{r}}dtdr - dr^2 - r^2\left(d\theta^2 + \sin^2\theta d\varphi^2\right).$$

(46)

The self-conjugate Hamiltonian $H_\eta$ is

$$H_\eta = \gamma^0 m - i\gamma^0\left\{\gamma^1\left(\frac{\partial}{\partial r} + \frac{1}{r}\right) + \gamma^2 \frac{1}{r}\left(\frac{\partial}{\partial \theta} + \frac{1}{2}\text{ctg}\theta\right) + \gamma^3 \frac{1}{r\sin\theta}\frac{\partial}{\partial \varphi}\right\} +$$
$$+i\sqrt{\frac{r_0}{r}}\left(\frac{\partial}{\partial r} + \frac{3}{4}\frac{1}{r}\right).$$

(47)

Here, similarly to the previous solution, the condition (45) should be fulfilled.

### 3.6 Finkelstein-Lemaitre metric

$$ds^2 = dt^2 - \frac{dr^2}{f_{F-L}^{2/3}} - f_{F-L}^{4/3}r^2\left(d\theta^2 + \sin^2\theta d\varphi^2\right),$$

$$f_{F-L} = \frac{3}{2r_0}(r-t).$$

(48)

The self-conjugate Hamiltonian $H_\eta$ is

$$H_\eta = \gamma^0 m - i\gamma^0\gamma^1 f_{F-L}^{1/3}\left(\frac{\partial}{\partial r} + \frac{1}{r}\right) - i\gamma^0\gamma^2 \frac{1}{f_{F-L}^{2/3}} \frac{1}{r_0}\left(\frac{\partial}{\partial \theta} + \frac{1}{2}\operatorname{ctg}\theta\right) -$$
$$-i\gamma^0\gamma^3 \frac{1}{f_{F-L}^{2/3}} \frac{1}{r_0 \sin\theta} \frac{\partial}{\partial \varphi} - \frac{i}{2}\gamma^0\gamma^1 \frac{\partial f_{F-L}^{1/3}}{\partial r}.$$
(49)

### 3.7 Kruskal metric

$$ds^2 = dt^2 - f_K dr^2 - \frac{1}{4} f_K r_0^2 \left(d\theta^2 + \sin^2\theta d\varphi^2\right);$$
$$\frac{t}{r_0} = \frac{1}{2}\left(1 + \frac{r^2}{r_0^2}\right)^{3/2}(\pi - \chi + \sin\chi);$$
(50)
$$f_K = \left(1 + \frac{r^2}{r_0^2}\right)(1 - \cos\chi)^2.$$

The self-conjugate Hamiltonian $H_\eta$ is

$$H_\eta = \gamma^0 m - i\gamma^0\gamma^1 \frac{1}{\sqrt{f_K}}\left(\frac{\partial}{\partial r} + \frac{1}{r}\right) - i\gamma^0\gamma^2 \frac{2}{\sqrt{f_K}} \frac{1}{r_0}\left(\frac{\partial}{\partial \theta} + \frac{1}{2}\operatorname{ctg}\theta\right) -$$
$$-i\gamma^0\gamma^3 \frac{2}{\sqrt{f_K}} \frac{1}{r_0 \sin\theta} \frac{\partial}{\partial \varphi} - \frac{i}{2}\gamma^0\gamma^1 \frac{\partial f_K^{-1/2}}{\partial r}.$$
(51)

For the Finkelstein-Lemaitre and Kruskal solutions, the time coordinate coincides with the proper time. The domain of the wave functions is the entire space $(r,\theta,\varphi)$. The Hamiltonians $H_\eta$ for these solutions are explicitly time-dependent, and stationary bound states cannot therefore exist.

We further analyze the possibility of existence of bound states for the metrics with stationary Hamiltonians (32), (36), (41), (44), (47) with the wave function represented as

$$\psi_\eta(\mathbf{r},t) = e^{-iEt}\psi_\eta(\mathbf{r}).$$
(52)

In (52) $E$ - energy of Dirac particle.

### 3.8 Separation of variables

The Dirac equation with stationary Hamiltonians (32), (36), (41), (44), (47) allows for the separation of variables, if the bispinor $\psi(\mathbf{r}) = \psi(r,\theta,\varphi)$ is defined as

$$\psi_\eta(r,\theta,\varphi) = \begin{pmatrix} F(r)\cdot\xi(\theta) \\ -iG(r)\cdot\sigma^3\xi(\theta) \end{pmatrix} e^{im_\varphi\varphi}$$
(53)

and the following equation is used (see, e.g., [26])

$$\left[-\sigma^2\left(\frac{\partial}{\partial\theta}+\frac{1}{2}\mathrm{ctg}\,\theta\right)+i\sigma^1 m_\varphi \frac{1}{\sin\theta}\right]\xi(\theta)=i\kappa\xi(\theta). \tag{54}$$

In order to receive Eq. (54) we made an equivalent replacement of matrices in Hamiltonians (32), (36), (41), (44), (47):

$$\gamma^1 \to \gamma^3,\ \gamma^3 \to \gamma^2,\ \gamma^2 \to \gamma^1 \tag{55}$$

In (53), (54), $\xi(\theta)$ are spherical harmonics for spin ½, $\sigma^i$ are two-dimensional Pauli matrices, $m_\varphi$ is the magnetic quantum number, and $\kappa$ is the quantum number of the Dirac equation:

$$\kappa = \pm 1, \pm 2 \ldots = \begin{cases} -(l+1),\ j = l + \tfrac{1}{2} \\ l,\quad j = l - \tfrac{1}{2} \end{cases}. \tag{56}$$

In (56), $j, l$ are the quantum numbers of the total and orbital momentum of a Dirac particle, respectively.

$\xi(\theta)$ can be represented as [27]

$$\xi(\theta) = \begin{pmatrix} {}_{-\tfrac{1}{2}}Y_{jm_\varphi}(\theta) \\ {}_{\tfrac{1}{2}}Y_{jm_\varphi}(\theta) \end{pmatrix} = (-1)^{m_\varphi + \tfrac{1}{2}}\sqrt{\frac{1}{4\pi}\frac{(j-m_\varphi)!}{(j+m_\varphi)!}}\begin{pmatrix} \cos\tfrac{\theta}{2} & \sin\tfrac{\theta}{2} \\ -\sin\tfrac{\theta}{2} & \cos\tfrac{\theta}{2} \end{pmatrix} \times$$
$$\times \begin{pmatrix} \left(\kappa - m_\varphi + \tfrac{1}{2}\right)\cdot P_l^{m_\varphi - \tfrac{1}{2}}(\theta) \\ P_l^{m_\varphi + \tfrac{1}{2}}(\theta) \end{pmatrix}. \tag{57}^{3}$$

In (57), $P_l^{m_\varphi \pm \tfrac{1}{2}}(\theta)$ are Legendre polynomials.

As a result of the separation of variables, for the above solutions, we obtain a system of equations for the real radial functions $F(r), G(r)$. In the following these equations are written in the dimensionless variables $\varepsilon = \dfrac{E}{m}, \rho = \dfrac{r}{l_c}, 2\alpha = \dfrac{r_0}{l_c}$, where $l_c = \dfrac{\hbar}{mc}$ is the Compton wavelength of Dirac particle.

---

[3] In (57) $\begin{pmatrix} \cos\tfrac{\theta}{2} & \sin\tfrac{\theta}{2} \\ -\sin\tfrac{\theta}{2} & \cos\tfrac{\theta}{2} \end{pmatrix}$ is a 2x2 matrix.

## 3.9 Equations, asymptotics and boundary conditions for radial wave functions. Hermiticity of Hamiltonians

### 3.9.1. Schwarzschild solution in the $(t,r,\theta,\varphi)$ coordinates [1], in isotropic coordinates [2] and in harmonic coordinates [3]

Systems of equations for the real radial functions are written as follows:

For the metric (31) and Hamiltonian (32)

$$f_s \frac{dF_s}{d\rho} + \left(\frac{1+\kappa\sqrt{f_s}}{\rho} - \frac{\alpha}{\rho^2}\right)F_s - \left(\varepsilon + \sqrt{f_s}\right)G_s = 0,$$

$$f_s \frac{dG_s}{d\rho} + \left(\frac{1-\kappa\sqrt{f_s}}{\rho} - \frac{\alpha}{\rho^2}\right)G_s + \left(\varepsilon - \sqrt{f_s}\right)F_s = 0.$$

(58)

For the metric (33) and Hamiltonian (36)

$$F_{Ob}\frac{dF_{is}}{d\rho} + F_{Ob}\frac{1+\kappa}{\rho}F_{is} + \frac{1}{2}\frac{dF_{Ob}}{d\rho}F_{is} - (\varepsilon + V)G_{is} = 0,$$

$$F_{Ob}\frac{dG_{is}}{d\rho} + F_{Ob}\frac{1-\kappa}{\rho}G_{is} + \frac{1}{2}\frac{dF_{Ob}}{d\rho}G_{is} + (\varepsilon - V)F_{is} = 0.$$

(59)

For the metric (39) and Hamiltonian (41)

$$F_g\frac{dF_{gr}}{d\rho} + F_g\frac{F_{gr}}{\rho} + \frac{\sqrt{F_g}}{\left(1+\frac{\alpha}{\rho}\right)\rho}\kappa F_{gr} + \frac{1}{2}\frac{dF_g}{d\rho}F_{gr} - \left(\varepsilon + \sqrt{F_g}\right)G_{gr} = 0,$$

$$F_g\frac{dG_{gr}}{d\rho} + F_g\frac{G_{gr}}{\rho} - \frac{\sqrt{F_g}}{\left(1+\frac{\alpha}{\rho}\right)\rho}\kappa G_{gr} + \frac{1}{2}\frac{dF_g}{d\rho}G_{gr} + \left(\varepsilon - \sqrt{F_g}\right)F_{gr} = 0.$$

(60)

Radii of the "event horizons" for the metrics (31), (33), (39) are $r_0$, $\frac{r_0}{4}$, $\frac{r_0}{2}$, respectively. If the condition $g_{00} > 0$ is fulfilled, the domains of the functions $F(\rho), G(\rho)$ are the intervals $\rho \in (2\alpha, \infty)$, $\rho \in \left(\frac{\alpha}{2}, \infty\right)$, $\rho \in (\alpha, \infty)$, respectively. The wave functions must be zero on and under the "event horizons".

Let us consider the asymptotics of solutions of (58) - (60).
For $\rho \to \infty$, for each system of equations the leading terms of asymptotics equal

$$F = C_1 e^{-\rho\sqrt{1-\varepsilon^2}} + C_2 e^{\rho\sqrt{1-\varepsilon^2}},$$

$$G = \sqrt{\frac{1-\varepsilon}{1+\varepsilon}}\left(-C_1 e^{-\rho\sqrt{1-\varepsilon^2}} + C_2 e^{\rho\sqrt{1-\varepsilon^2}}\right).$$

(61)

In order to provide the finite motion of Dirac particles, we must use only exponentially decreasing solutions (61), i.e. in this case $C_2 = 0$.

The behavior of the wave functions close to the "event horizons" has the following form: for Eqs. (58) with $\rho \to 2\alpha$ $(r \to r_0)$

$$F_s = \frac{A_s}{\sqrt{\rho - 2\alpha}} \sin(2\alpha\varepsilon \ln(\rho - 2\alpha) + \varphi_s),$$
$$G_s = \frac{A_s}{\sqrt{\rho - 2\alpha}} \cos(2\alpha\varepsilon \ln(\rho - 2\alpha) + \varphi_s); \tag{62}$$

for Eqs. (59) with $\rho \to \frac{\alpha}{2}$ $\left(r \to \frac{r_0}{4}\right)$

$$F_{is} = \frac{A_{is}}{\rho F_{Ob}^{1/2}} \sin\left(8\alpha\varepsilon \ln\left(\rho - \frac{\alpha}{2}\right) + \varphi_{is}\right),$$
$$G_{is} = \frac{A_{is}}{\rho F_{Ob}^{1/2}} \cos\left(8\alpha\varepsilon \ln\left(\rho - \frac{\alpha}{2}\right) + \varphi_{is}\right); \tag{63}$$

for Eqs. (60) with $\rho \to \alpha$ $\left(r \to \frac{r_0}{2}\right)$

$$F_{gr} = \frac{A_{gr}}{\rho F_g^{1/2}} \sin(2\alpha\varepsilon \ln(\rho - \alpha) + \varphi_{gr}),$$
$$G_{gr} = \frac{A_{gr}}{\rho F_g^{1/2}} \cos(2\alpha\varepsilon \ln(\rho - \alpha) + \varphi_{gr}). \tag{64}$$

In (62) - (64), $A_s, \varphi_s; A_{is}, \varphi_{is}; A_{gr}, \varphi_{gr}$ are integration constants.

The oscillating functions $F$ and $G$ in (62) - (64) are ill-defined at the "event horizons", but they are quadratically integrable functions at $\rho_{\min} \neq 2\alpha$, at $\rho_{\min} \neq \frac{\alpha}{2}$, at $\rho_{\min} \neq \alpha$, respectively.

The Hamiltonians (32), (36), (41) are Hermitian over the whole domains of $\rho$.

We can show this using the general Hermiticity condition for Dirac Hamiltonians in external gravitational fields proven in [10].

$$\oint ds_k \left(\sqrt{-g} j^k\right) + \int d^3x \sqrt{-g} \left[\psi^+ \gamma^0 \left(\gamma^0_{,0} + \begin{pmatrix} 0 \\ 00 \end{pmatrix} \gamma^0 \right) \psi + \begin{pmatrix} k \\ k0 \end{pmatrix} j^0 \right] = 0. \tag{65}$$

For the time-independent Hamiltonians (32), (36), (41) $\gamma^0_{,0} \equiv \frac{\partial \gamma^0}{\partial x^0} = 0$, Christoffel symbols $\begin{pmatrix} 0 \\ 00 \end{pmatrix}, \begin{pmatrix} k \\ k0 \end{pmatrix} = 0$ and condition (64) reduces to

$$\oint ds_k \left(\sqrt{-g}\, j^k\right) = 0. \tag{66}$$

Considering the equivalent replacement $\gamma$ - matrix (55) the components of the current density of Dirac particles are equal to:

1. For the metric (31) and Hamiltonian (32)

$$j_s^r = \psi_\eta^+ f_s \gamma^0 \gamma^3 \psi_\eta, \tag{67}$$

$$j_s^\theta = \psi_\eta^+ f_s^{1/2} \gamma^0 \gamma^1 \psi_\eta, \tag{68}$$

$$j_s^\varphi = \psi_\eta^+ f_s^{1/2} \gamma^0 \gamma^2 \psi_\eta. \tag{69}$$

For the Hamiltonian (32) the wave function (53) can be written as

$$\psi_s(\rho,\theta,\varphi) = \frac{1}{\rho}\frac{1}{\sqrt{f_s}}\begin{pmatrix} f_s(\rho)\xi(\theta) \\ -ig_s(\rho)\sigma^3\xi(\theta) \end{pmatrix} e^{im_\varphi \varphi}. \tag{70}$$

Considering (67) - (70) and the explicit form of the spherical harmonics (57),

$$j_s^r = \frac{i}{\rho^2} f_s(\rho) g_s(\rho) \left[\xi^+(\theta)\left(\sigma^3\sigma^3 - \sigma^3\sigma^3\right)\xi(\theta)\right] = 0, \tag{71}$$

$$j_s^\theta = -\frac{2}{\rho^3 f_s^{1/2}} f_s(\rho) g_s(\rho) \left[\xi^+(\theta)\sigma^2\xi(\theta)\right] = 0, \tag{72}$$

$$j_s^\varphi = \frac{2}{\rho^2 f_s^{1/2}} f_s(\rho) g_s(\rho) \left[\xi^+(\theta)\sigma^1\xi(\theta)\right] \neq 0. \tag{73}$$

The $\varphi$-component of the current is non-zero and it grows infinitely as $\rho \to 2\alpha$ $(r \to r_0)$.

2. For the metric (33) and Hamiltonian (36)

$$j_{is}^{r,\theta,\varphi} = \psi_\eta^+ F_{Ob} \gamma^0 \left(\gamma^{3,1,2}\right) \psi_\eta. \tag{74}$$

For the Hamiltonian (36) the wave function (53) can be written as

$$\psi_{is}(\rho,\theta,\varphi) = \frac{1}{\rho F_{Ob}^{1/2}}\begin{pmatrix} f_{is}(\rho)\cdot\xi(\theta) \\ -ig_{is}(\rho)\cdot\sigma^3\xi(\theta) \end{pmatrix} e^{im_\varphi \varphi}. \tag{75}$$

Considering (74) – (75),

$$j_{is}^r = \frac{i}{\rho^2} f_{is}(\rho) g_{is}(\rho) \left[\xi^+(\theta)\left(\sigma^3\sigma^3 - \sigma^3\sigma^3\right)\xi(\theta)\right] = 0, \tag{76}$$

$$j_{is}^\theta = -\frac{2}{\rho^2} f_{is}(\rho) g_{is}(\rho) \left[\xi^+(\theta)\sigma^2\xi(\theta)\right] = 0, \tag{77}$$

$$j_{is}^\varphi = \frac{2}{\rho^2} f_{is}(\rho) g_{is}(\rho) \left[\xi^+(\theta)\sigma^1\xi(\theta)\right] \neq 0. \tag{78}$$

3. For the metric (39) and Hamiltonian (41)

$$j_{gr}^r = \psi_\eta^+ F_g \gamma^0 \gamma^3 \psi_\eta$$

$$j_{gr}^{\theta,\varphi} = \psi_\eta^+ \frac{F_g^{1/2}}{\left(1+\dfrac{\alpha}{\rho}\right)} \gamma^0 \left(\gamma^{1,2}\right) \psi_\eta. \tag{79}$$

For the Hamiltonian (41) the wave function (53) can be written as

$$\psi_{gr}(\rho,\theta,\varphi) = \frac{1}{\rho F_g^{1/2}} \begin{pmatrix} f_{gr}(\rho)\, \xi(\theta) \\ -ig_{gr}(\rho)\, \sigma^3 \xi(\theta) \end{pmatrix} e^{im_\varphi \varphi}. \tag{80}$$

Considering (79) – (80),

$$j_{gr}^r = \frac{i}{\rho^2} f_{gr}(\rho) g_{gr}(\rho) \left[\xi^+(\theta)\left(\sigma^3 \sigma^3 - \sigma^3 \sigma^3\right)\xi(\theta)\right] = 0, \tag{81}$$

$$j_{gr}^\theta = -\frac{2}{\rho^2} \frac{1}{F_g^{1/2}\left(1+\dfrac{\alpha}{\rho}\right)} f_{gr}(\rho) g_{gr}(\rho) \left[\xi^+(\theta)\sigma^2 \xi(\theta)\right] = 0, \tag{82}$$

$$j_{gr}^\varphi = \frac{i}{\rho^2 F_g^{1/2}\left(1+\dfrac{\alpha}{\rho}\right)} f_{gr}(\rho) g_{gr}(\rho) \left[\xi^+(\theta)\sigma^1 \xi(\theta)\right] \neq 0. \tag{83}$$

The $\varphi$-component differs from zero and grows infinitely as $\rho \to \alpha$ $\left(r \to \dfrac{r_0}{2}\right)$.

For a centrally symmetric Schwarzschild field, the Hermiticity condition (65) can be written as

$$4\pi \rho^2 j^r (\rho \to \infty) + 4\pi \rho^2 j^r (\rho \to 2\alpha) = 0. \tag{84}$$

The equalities (71), (76), (81) show that the condition (84) is fulfilled for all the three metrics under consideration.

Thus, if we introduce physically reasonable boundary conditions close to the "event horizon", the systems of equations (58) - (60) will possess stationary real energy spectra of bound states of spin-half particles.

In [14], a constraint of the $\varphi$-component of the Dirac current close to the "event horizon" is selected as such boundary condition for the metric (31).

As the simplest constraint the condition is suggested

$$f_s(\rho) g_s(\rho)\big|_{\rho \to 2\alpha} = 0. \tag{85}$$

### 3.9.2. Eddington-Finkelstein [4], [7] and Painlevé-Gullstrand [8], [9] metrics

Systems of equations for the radial wave functions $F(\rho), G(\rho)$ are written as follows:

For the metric (43) and Hamiltonian (44)

$$\left(1-\frac{2\alpha}{\rho}\right)\frac{dF_{E-F}}{d\rho} + \left(\frac{1-\frac{2\alpha}{\rho}}{\rho} + \frac{\frac{\kappa}{\sqrt{f_{E-F}}}}{\rho} + \frac{1}{f_{E-F}}\frac{\alpha}{\rho^2}\right)F_{E-F} - i\frac{2\alpha}{\rho}\left(\varepsilon - \frac{1}{\sqrt{f_{E-F}}}\right)F_{E-F} +$$

$$+\frac{1}{f_{E-F}}\frac{\alpha}{\rho^2}G_{E-F} - \frac{2\alpha}{\rho^2}\frac{\kappa}{\sqrt{f_{E-F}}}G_{E-F} - i\left(\varepsilon + \frac{1}{\sqrt{f_{E-F}}}\right)G_{E-F} = 0$$

(86)

$$\left(1-\frac{2\alpha}{\rho}\right)\frac{dG_{E-F}}{d\rho} + \left(\frac{1-\frac{2\alpha}{\rho}}{\rho} - \frac{\frac{\kappa}{\sqrt{f_{E-F}}}}{\rho} + \frac{1}{f_{E-F}}\frac{\alpha}{\rho^2}\right)G_{E-F} - i\frac{2\alpha}{\rho}\left(\varepsilon + \frac{1}{\sqrt{f_{E-F}}}\right)G_{E-F} +$$

$$+\frac{1}{f_{E-F}}\frac{\alpha}{\rho^2}F_{E-F} + \frac{2\alpha}{\rho^2}\frac{\kappa}{\sqrt{f_{E-F}}}F_{E-F} - i\left(\varepsilon - \frac{1}{\sqrt{f_{E-F}}}\right)F_{E-F} = 0.$$

In (86), $f_{E-F} = 1 + \frac{2\alpha}{\rho}$.

For the metric (46) and Hamiltonian (47)

$$\left(1-\frac{2\alpha}{\rho}\right)\frac{dF_{P-G}}{d\rho} + \left\{\frac{1+\kappa}{\rho} - \frac{3}{4}\frac{2\alpha}{\rho^2} + i(1-\varepsilon)\sqrt{\frac{2\alpha}{\rho}}\right\}F_{P-G} - \left\{1+\varepsilon + i\sqrt{\frac{2\alpha}{\rho}}\frac{\frac{1}{4}-\kappa}{\rho}\right\}G_{P-G} = 0$$

(87)

$$\left(1-\frac{2\alpha}{\rho}\right)\frac{dG_{P-G}}{d\rho} + \left\{\frac{1-\kappa}{\rho} - \frac{3}{4}\frac{2\alpha}{\rho^2} - i(1+\varepsilon)\sqrt{\frac{2\alpha}{\rho}}\right\}G_{P-G} - \left\{1-\varepsilon - i\sqrt{\frac{2\alpha}{\rho}}\frac{\frac{1}{4}+\kappa}{\rho}\right\}F_{P-G} = 0.$$

Asymptotics of the functions $F(\rho), G(\rho)$ with $\rho \to \infty$ coincides with asymptotics of (61) for the Schwarzschild metrics (31), (33), (39).

The gravitational radius for Eqs. (86), (87) is a special point, at which a certain relationship between the functions $F(\rho = 2\alpha)$ and $G(\rho = 2\alpha)$ is reached. In [28], it was established numerically for the Hamiltonian (47) that the functions $F_{P-G}, G_{P-G}$ are smooth close to the "event horizons".

With such a behavior of the wave functions, the initial Hamiltonians (44), (47) are non-Hermitian,

$$(\psi, H\varphi) \neq (H\psi, \varphi).$$

(88)

We rewrite the condition (65) in the following form

$$\oint ds_k \left( \sqrt{-g}\, j^k \right) + \int d^3x \sqrt{-g}\, \psi^+ \gamma^{\underline{0}} \left( \gamma^{\underline{0}}_{,0} + \begin{pmatrix} 0 \\ 00 \end{pmatrix} \gamma^{\underline{0}} + \begin{pmatrix} k \\ k0 \end{pmatrix} \gamma^{\underline{0}} \right) \psi = 0. \tag{89}$$

In (89), like previously, we use summation over spatial index $k = 1, 2, 3$.

For the Eddington-Finkelstein and Painlevé-Gullstrand metrics $\gamma^{\underline{0}}_{,0} \equiv \dfrac{\partial \gamma^{\underline{0}}}{\partial x^0} = 0$,

$\begin{pmatrix} 0 \\ 00 \end{pmatrix} + \begin{pmatrix} 1 \\ 10 \end{pmatrix} = 0$, $\begin{pmatrix} 2 \\ 20 \end{pmatrix} = \begin{pmatrix} 3 \\ 30 \end{pmatrix} = 0$. The condition (89) reduces to the condition (66).

In this case the relation (84) should be fulfilled

$$4\pi\rho^2 j^r \left( \rho \to \infty \right) + 4\pi\rho^2 j^r \left( \rho \to 2\alpha \right) = 0. \tag{90}$$

Considering the equivalent replacement $\gamma$ - matrices (55) at variables separation for metric (43) and Hamiltonian (44)

$$j^r_{E-F} = \psi^+_\eta \left( 1 + \frac{2\alpha}{\rho} \right)^{-1} \left( -\frac{2\alpha}{\rho} + \gamma^{\underline{0}} \gamma^{\underline{3}} \right) \psi_\eta. \tag{91}$$

Considering the asymptotics of (61) with $C_2 = 0$, in the Hermiticity condition (90), the expression $4\pi\rho^2 j^r$ tends to zero.

For $\rho \to 2\alpha$, considering (53) the expression $4\pi\rho^2 j^r$ is finite.

$$\left( 4\pi\rho^2 j^r \right)\bigg|_{\rho \to 2\alpha} = \left\{ 4\pi\rho^2 \left[ -\frac{2\alpha}{\rho} \frac{1}{1 + \dfrac{2\alpha}{\rho}} \left( F^2_{E-F}(\rho) + G^2_{E-F}(\rho) \right) \left[ \xi^+(\theta) \xi(\theta) \right] + \right. \right.$$
$$\left. \left. + \frac{i}{1 + \dfrac{2\alpha}{\rho}} F_{E-F}(\rho) G_{E-F}(\rho) \left[ \xi^+(\theta) \left( \sigma^3 \sigma^3 - \sigma^3 \sigma^3 \right) \xi(\theta) \right] \right] \right\} \bigg|_{\rho \to 2\alpha}. \tag{92}$$

In (92) the first summand is finite. The Hermiticity condition (90) for the Eddington-Finkelstein solution is not satisfied.

Stationary bound states of Dirac particles are absent in this case. Only decaying with time complex energy levels can exist.

For the Painlevé-Gullstrand metric (46) and Hamiltonian (47) considering the equivalent replacement $\gamma$ - matrices (55)

$$j^r_{P-G} = \psi^+_\eta \left( -\sqrt{\frac{2\alpha}{\rho}} + \gamma^{\underline{0}} \gamma^{\underline{3}} \right) \psi_\eta. \tag{93}$$

The Hermiticity condition (90) is not fulfilled because of the finite first summand in (93) at $\rho \to 2\alpha$.

Thus, for both metrics (43), (46), stationary bound states of Dirac particles cannot exist. However, because of the fulfillment of the Hilbert condition $g_{00} > 0$ the domain of the wave functions is an interval $\rho \in (2\alpha, \infty)$ and the "event horizon" for these metrics, as in the case of the metrics (31), (33), (39), is an infinitely high potential barrier.

### 4. Conclusions

As a result of the quantum-mechanical analysis we established that the metrics of a centrally symmetric gravitational field [1] - [9], which are exact solutions of the general relativity equations, can be divided into three groups:

1. For the classical Schwarzschild metric [1], the Schwarzschild metric in isotropic coordinates [2] and the Schwarzschild metric in harmonic coordinates [3], stationary bound states of spin-half particles can exist. As a result of the fulfillment of the Hilbert condition $g_{00} > 0$, the "event horizon" becomes an infinitely high potential barrier for quantum-mechanical Dirac particles. For such black holes, the wave function under the "event horizon" is zero and Hawking radiation [17] is absent.

2. For the Eddington-Finkelstein [4], [7] and Painlevé-Gullstrand [8], [9] metrics, because the Hamiltonians are non-Hermitian, the non-decaying bound states of Dirac particles cannot exist. However, the fulfillment of the Hilbert condition $g_{00} > 0$ again makes the "event horizons" insurmountable for spin-half particles. For these solutions, the wave functions under the "event horizon" are zero, and such black holes do not radiate through the Hawking mechanism.

3. For the Finkelstein-Lemaitre [4] and Kruskal [5], [6] metrics, the Dirac Hamiltonians explicitly depend time, and stationary bound states cannot exist. The metrics do not impose any constraints on the domain of the wave functions, and for these metrics, Hawking evaporation of black holes is therefore possible.

It is evident that the three groups of metrics are physically non-equivalent in the quantum-mechanical framework. However, all the metrics are general relativity solutions and they can be implemented during the post-inflationary period of the universe's expansion.

The results of this work can lead to revisiting some concepts of the standard cosmological model related to the evolution of the universe and interaction of collapsars with surrounding matter.

**Acknowledgement**

We thank A.L. Novoselova for her significant technical help in the preparation of the paper.

**References**

[1]   K.Schwarzschild. Sitzber. Deut. Akad. Wiss. Berlin, 189-196 (1916).

[2] A.S.Eddington. The Mathematical Theory of Relativity (Cambridge University Press, 1924).
       Yu.N. Obukhov. Phys. Rev. Lett. 86, 192 (2001): Forschr. Phys. 50, 711 (2002).

[3] A.A. Logunov, M.A. Mestvirishvili. Relativistic gravitation theory. M.: Nauka (1989) (in Russian).

[4] D. Finkelstein. Phys. Rev. 110, 965 (1958).

[5] M. Kruskal. Phys. Rev. 119, 1743 (1960).

[6] I.D.Novikov. AJ 40, 772 (1963) (in Russian).

[7] A.S.Eddington. Nature 113, 192 (1924).

[8] P.Painleve. C.R.Acad. Sci. (Paris) 173, 677 (1921).

[9] A.Gullstrand. Arkiv. Mat. Astron. Fys. 16, 1 (1922).

[10]   M.V.Gorbatenko, V.P.Neznamov. Phys. Rev. D 82, 104056 (2010); arxiv: 1007.4631 (gr-qc).

[11]   M.V.Gorbatenko, V.P.Neznamov. Phys. Rev. D 83, 105002 (2011); arxiv: 1102.4067v1 (gr-qc).

[12]   M.V.Gorbatenko, V.P.Neznamov. Arxiv: 1107.0844 (gr-qc).

[13]   M.V.Gorbatenko, V.P.Neznamov. Arxiv: 1205.4348 (gr-qc).

[14]   M.V.Vronsky, M.V.Gorbatenko, N.S.Kolesnikov, V.P.Neznamov, E.V.Popov, I.I.Safronov. Arxiv: 1301.7595 (gr-qc).

[15]   D.Hilbert. Math. Ann., 1924. Bd. 92. S. 1-32.

[16]   L.D.Landau and E.M.Lifshitz, The Classical theory of Fields (Pergamon Press, Oxford, 1975).

[17]   W.Hawking. Common. Math. Phys. 43, 199-220 (1975).

[18]   M.V.Gorbatenko, V.P.Neznamov. Arxiv: 1302.2557 (gr-qc).

[19]   M.V.Gorbatenko, V.P.Neznamov. Arxiv: 1303.1127 (gr-qc).

[20]   H.Reissner, Ann. Phys. 50, 106 (1916).

[21]   C.Nordström, Proc.K.Akad.Wet.Amsterdam 20, 1238 (1918).

[22] R.P.Kerr. Phys. Rev. Lett. 11, 237 (1963).

[23] E.T.Newman, E.Couch, K.Chinnapared, A.Exton, A.Prakash and R.Torrence, J. Math. Phys. 6, 918 (1965).

[24] V.A.Fok. Theory of space, time and gravitation. Moscow, GITTL, 1955 [in Russian].

[25] J.L.Synge. Relativity: the general theory. North-Holland Publishing Company. Amsterdam, 1960.

[26] D.R.Brill, J.A.Wheeler. Rev. of Modern Physics, 29, 465-479 (1957).

[27] S.R.Dolan. Trinity Hall and Astrophysics Group, Cavendish Laboratory. Dissertation, 2006.

[28] A.Lasenby, C.Doran, J.Pritchard, A.Caceres and S.Dolan. Phys. Rev. D 72, 105014 (2005).